\documentclass[aps,prb,superscriptaddress,twocolumn]{revtex4-2}
\usepackage[margin=1.5cm]{geometry}
\usepackage{graphicx}

\usepackage[caption=false]{subfig}

\usepackage[T1]{fontenc}

\usepackage{lineno,hyperref}
\usepackage{hyperref}
\usepackage{amsmath}

\makeatletter

\newenvironment{figurehere}
  {\def\@captype{figure}}
  {}
\makeatother

\begin{document}


\title{Photoemission signature of momentum-dependent hybridization in CeCoIn$_5$}


\author{R. Kurleto}
\affiliation{Marian Smoluchowski Institute of Physics, Jagiellonian University, {\L}ojasiewicza 11, 30-348 Krak{\'o}w, Poland}
\author{M. Fidrysiak}
\affiliation{Institute of Theoretical Physics, Jagiellonian University, {\L}ojasiewicza 11, 30-348 Krak{\'o}w, Poland}

\author{L.~Nicola\"{i}}
\affiliation{New Technologies-Research Center, University of West Bohemia, Univerzitn{\'i­} 8, 306 14 Pilsen, Czech Republic}

\author{J.~Min\'{a}r}
\affiliation{New Technologies-Research Center, University of West Bohemia, Univerzitn{\'i­} 8, 306 14 Pilsen, Czech Republic}

\author{M.~Rosmus}
\affiliation{Marian Smoluchowski Institute of Physics, Jagiellonian University, {\L}ojasiewicza 11, 30-348 Krak{\'o}w, Poland}
\affiliation{Solaris National Synchrotron Radiation Centre, Jagiellonian University, Czerwone Maki 98, 30-392 Krak{\'o}w, Poland}

\author{{\L}.~Walczak}
\affiliation{PREVAC sp. z o.o., Raciborska 61, PL-44362 Rog\'ow, Poland}

\author{A.~Tejeda}
\affiliation{Laboratoire de Physique des Solides, CNRS, Universit\'e Paris-Sud, Universit\'e Paris-Saclay, 91405 Orsay, France}

\author{J.~E.~Rault}
\affiliation{Synchrotron-SOLEIL, Saint-Aubin, BP48, F91192 Gif sur Yvette Cedex, France}

\author{F.~Bertran}
\affiliation{Synchrotron-SOLEIL, Saint-Aubin, BP48, F91192 Gif sur Yvette Cedex, France}

\author{A.~P.~K{\k{a}}dzielawa}
\affiliation{IT4Innovations, V\v{S}B - Technical University of Ostrava, 17. listopadu 2172/15, 708 00 Ostrava-Poruba, Czech Republic}
\affiliation{Institute of Theoretical Physics, Jagiellonian University, {\L}ojasiewicza 11, 30-348 Krak{\'o}w, Poland}

\author{D.~Legut}
\affiliation{IT4Innovations, V\v{S}B - Technical University of Ostrava, 17. listopadu 2172/15, 708 00 Ostrava-Poruba, Czech Republic}

\author{D.~Gnida}
\affiliation{Institute of Low Temperature and Structure Research,
Polish Academy of Sciences, P.O. Box 1410, 50-950 Wroc{\l}aw, Poland}

\author{D.~Kaczorowski}
\affiliation{Institute of Low Temperature and Structure Research,
Polish Academy of Sciences, P.O. Box 1410, 50-950 Wroc{\l}aw, Poland}

\author{K. Kissner}
\affiliation{Experimentelle Physik VII and W\"urzburg-Dresden Cluster of Excellence ct.qmat, Universit\"at W\"urzburg, Am Hubland, D-97074 W\"urzburg, Germany}

\author{F. Reinert}
\affiliation{Experimentelle Physik VII and W\"urzburg-Dresden Cluster of Excellence ct.qmat, Universit\"at W\"urzburg, Am Hubland, D-97074 W\"urzburg, Germany}


\author{J.~Spa{\l}ek}
\affiliation{Institute of Theoretical Physics, Jagiellonian University, {\L}ojasiewicza 11, 30-348 Krak{\'o}w, Poland}

\author{P.~Starowicz}
\affiliation{Marian Smoluchowski Institute of Physics, Jagiellonian University, {\L}ojasiewicza 11, 30-348 Krak{\'o}w, Poland}
\email[]{pawel.starowicz@uj.edu.pl}


\date{\today}

\begin{abstract}
Hybridization between $f$ electrons and conduction bands ($c$-$f$ hybridization) is a driving force for many unusual phenomena. To provide insight into it, systematic studies of CeCoIn$_5$ heavy fermion superconductor have been performed by angle-resolved photoemission spectroscopy (ARPES) in a large angular range at temperature of $T=6$~K. The used photon energy of 122 eV corresponds to Ce $4d$-$4f$ resonance. Calculations carried out with relativistic multiple scattering Korringa-Kohn-Rostoker method and one-step model of photoemission yielded realistic simulation of the ARPES spectra indicating that Ce-In surface termination prevails. Surface states, which have been identified in the calculations, contribute significantly to the spectra. Effects of the hybridization strongly depend on wave vector. They include a dispersion of heavy electrons and bands gaining $f$-electron character when approaching Fermi energy. We have also observed a considerable variation of $f$-electron spectral weight at $E_F$, which is normally determined by both matrix element effects and wave vector dependent $c$-$f$ hybridization. Fermi surface scans covering a few Brillouin zones revealed large matrix element effects. A symmetrization of experimental Fermi surface, which reduces matrix element contribution, yielded a specific variation of $4f$-electron enhanced spectral intensity at $E_F$ around $\bar{\Gamma}$ and \={M} points. Tight-binding approximation calculations for Ce-In plane provided the same universal distribution of $4f$-electron density for a range of values of the parameters used in the model. 


\end{abstract}


\maketitle

\section{Introduction}
{H}ybridization between $f$ electrons and conduction band ($c$-$f$ hybridization), together with strong $f$-$f$ correlations, is a source of peculiar electronic properties such as Kondo effect, heavy fermion or mixed valence state and can also play a role in quantum phase transitions and superconductivity~\citep{stewart,coleman2015heavy,si}. $c$-$f$ hybridization is introduced in theoretical description in a form of $V_{cf}$ hybridization parameter, which is present in a Hamiltonian of single impurity Anderson model~\citep{anderson1961} or periodic Anderson model~\citep{tahvildar}. A popular playground to investigate the mentioned phenomena is formed of systems containing Ce with $f$~electrons, which can be either localized or delocalized. The hybridization is often reflected in photoemission spectroscopy as a high intensity peak located near the Fermi energy ($E_F$) with strong contribution from $f$~electrons. This spectral feature is called Kondo or Abrikosov-Suhl resonance~\citep{allen,klein,sekiyama}. 

Early angle-resolved photoemission spectroscopy (ARPES) studies of the Kondo resonance for Ce systems already pointed out that its intensity varies with the wave vector, which is a fingerprint of anisotropic hybridization~\citep{garnier}. Later, it was observed that the Kondo peak is intense at Fermi vectors~\citep{danzenbacher}. Finally, it was shown that bands, which are strongly renormalized and characterized by high contribution from strongly correlated $f$ electrons, coexist with those weakly renormalized within a single heavy fermion system~\citep{koicz} and that hybridization changes with momentum~\citep{nakatani}. In fact, anisotropic structure of $V_{cf}$ influences Kondo resonance intensity variation along the Fermi surface (FS) and should contribute to a momentum dependence of quasiparticle weight and effective mass~\citep{ghaemi}. However, matrix element effects play also an important role here and have a serious impact on the spectral intensity. Although, the distribution of $f$-electron peak along FS can be investigated by ARPES, its theoretical modeling seems to be very complex. 

The effects introduced above can be conveniently studied on CeCoIn$_5$, which despite of relatively simple layered crystal structure, hosts many interesting physical phenomena~\citep{shimozawa,steglich}. These are superconductivity below $T_c=$~2.3~K~\citep{petrovic}, a heavy fermion state developed below coherence temperature equal to~45~K~\citep{jia} and characterized by the Sommerfeld coefficient of~$\gamma =$ 290~mJ/(mole$\times$K$^2$)~\citep{petrovic}, etc. Moreover, previous investigations proved that this system is suitable for ARPES studies, which have provided interesting results concerning FS, band structure, hybridization effects~\citep{koicz,dudy,chen1,jang,reber} and temperature dependence of heavy quasiparticle spectra~\citep{chen1,jang}.
%

In this paper, we report low-temperature ARPES studies for CeCoIn$_5$ performed in a large angular range covering a few Brillouin zones. One-step model of photoemission calculations provide a realistic simulation of the spectra including the matrix element effects. These shed light on experimental dispersions, as some of them appear to be related to the Ce-In surface. Various effects of hybridization are visible in the spectra. It is also possible to reduce geometrical effects on the spectral intensity related to matrix elements. It is done by a symmetrization of an experimental FS yielding $4f$-electron enhanced spectral distribution. Tight binding approximation (TBA) calculations for Ce-In layer reveal similar features of $4f$-electron weight distribution. 

\section{Material and methods}

Single crystalline samples of CeCoIn$_5$ have been obtained using the flux method~\citep{jia}. The samples have been cleaved in ultrahigh vacuum prior to the measurements (base pressure: 10$^{-11}$~mbar), exposing (001) plane. ARPES measurements have been realized at the Cassiopee beamline of Soleil synchrotron (Paris, France). All measurements have been performed at temperature equal to 6 K with Scienta R4000 photoelectron energy analyzer with parallel detection along a 30$^{\circ}$ large angular window. The range of the manipulator rotation angle~($\beta_{x}$) was up to approximately 50$^{\circ}$ to obtain FS mapping covering a few Brillouin zones (\hyperref[fig:jedynka1]{cf. Fig.~\ref*{fig:jedynka1}}). Spectra have been collected using photon energy of 122 eV (Ce $4d$-$4f$ resonant transition) and 88 eV. The polarization of the radiation was in the plane determined by incoming radiation, sample and the center of analyzer~(\hyperref[fig:jedynka1]{cf. Fig.~\ref*{fig:jedynka1}}). 

Self-consistent electronic structure calculations were performed within the \textit{ab initio} framework of spin-density functional theory using local density approximation. The electronic structure was calculated in a fully relativistic way by solving the corresponding Dirac equation using the relativistic multiple-scattering Korringa-Kohn-Rostocker (KKR) formalism as implemented in the SPR-KKR package~\citep{kkr}. The results of ground-state electronic structure calculations are represented here by the so-called Bloch spectral function (BSF), i.e., imaginary part of the Green's function. To understand ARPES results in great detail, we performed photoemission calculations based on the fully relativistic one-step model in its spin density matrix formulation~\citep{BME18}. The model accounts for all matrix element effects, the surface termination, and final state effects on the same level. All calculations have been performed for the semi-infinite surface geometry using various surface terminations and assuming real experimental geometry, photon energy, and light polarization.
  
The TBA model~\citep{maehira} based on the generalized Slater-Koster (SK) approach~\citep{slater} has been fitted to the experimental band structure. The five orbital TBA model for Ce-In planes comprises three doublets from Ce $4f_{\frac{5}{2}}$ manifold ($f_a$, $f_b$, $f_c$) together with two doublets ($p_a$, $p_b$; $j_z=\pm1$) from In $5p$ manifold. More details of the TBA calculations are given in~\hyperref[app:c]{Appendix~\ref*{app:c}}.

\section{Results and Discussion}

\begin{figure*}
{\centering\includegraphics[width=0.95\textwidth]{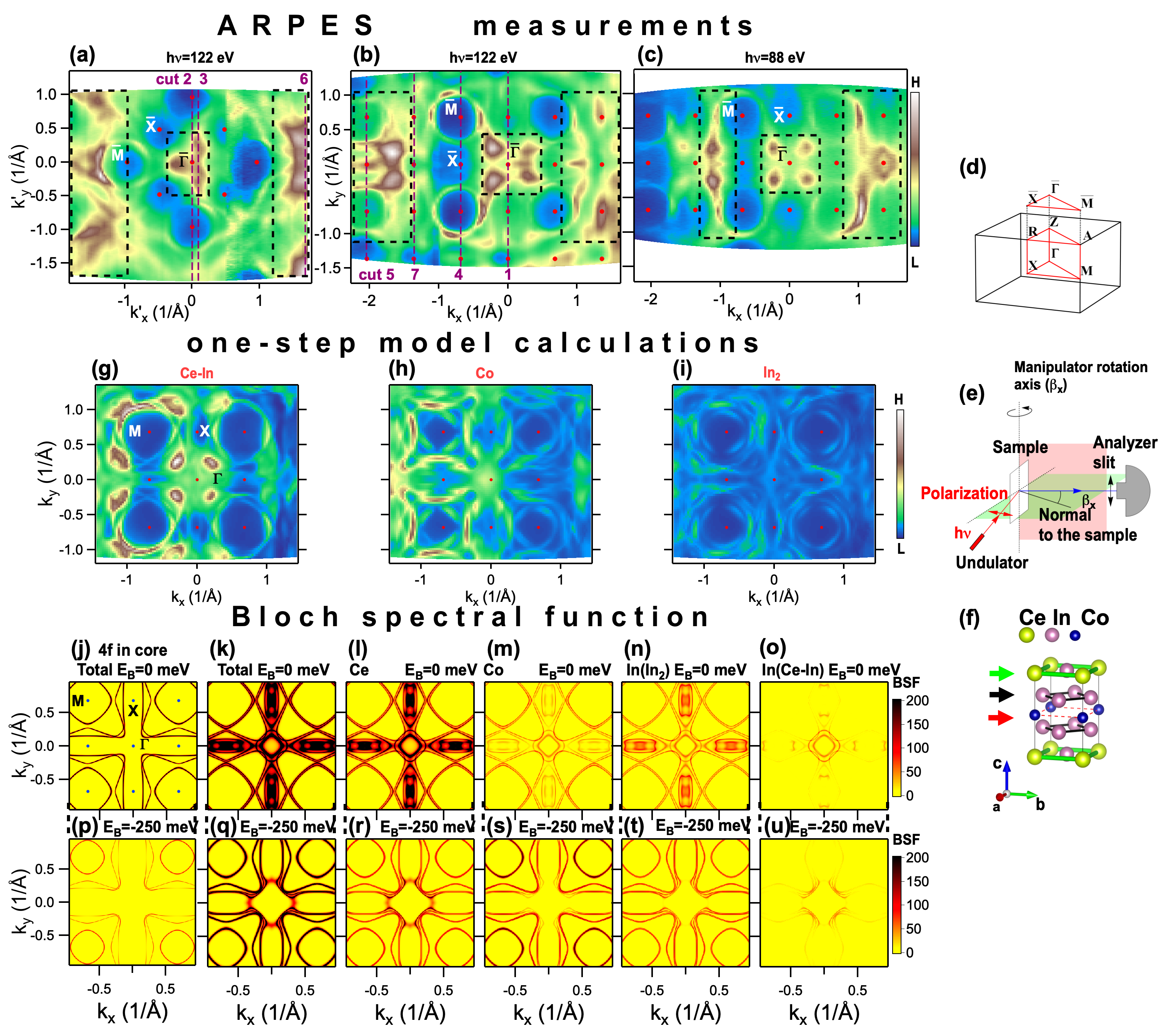}}

\caption{Fermi surface of CeCoIn$_5$.  Intensity maps along Fermi surface obtained by ARPES at $T$=6~K for photon energies of 122~eV (a,b) and 88~eV (c). Black dashed rectangles enclose FS regions with higher spectral intensity. (d)~Three-dimensional Brillouin zone and surface Brillouin zone with marked high symmetry points. (e)~Geometry of the experiment. Rotation of the manipulator ($\beta_{x}$) corresponds to the variation of  $k'_{x}$ or $k_{x}$ which depends on sample orientation. Linear horizontal polarization of the radiation is shown with red arrows. (f)~Crystal structure of CeCoIn$_5$.  One-step model based simulation of the ARPES spectra obtained for photon energy of 122~eV for Ce-In~(g), Co~(h), In$_2$~(i) surface terminations. Constant energy contours at 0~eV (j-o) and $-0.25$~eV (p-u) of Bloch spectral functions obtained in SPR-KKR calculations. These represent a topology of the Fermi surface. The total Bloch spectral function (k,q) is projected onto atomic wave functions of Ce (l,r), Co (m,s),  In from In$_2$ layers (n,t) and In from Ce-In planes (o,u). The total Bloch spectral function calculated in the absence of hybridization effects ($4f$ in core) is shown in (j) and (p).}
\label{fig:jedynka1}
\end{figure*}

The FS of CeCoIn$_5$ was studied by ARPES at low temperature~($T$=6~K) in a large angular range for two sample orientations~(\hyperref[fig:jedynka1]{cf. Fig.~\ref*{fig:jedynka1}~a-c}), namely with the analyzer slit along \={M}-$\bar{\Gamma}$-\={M}~(\hyperref[fig:jedynka1]{Fig.~\ref*{fig:jedynka1}~a}) and \={X}-$\bar{\Gamma}$-\={X}~(\hyperref[fig:jedynka1]{Fig.~\ref*{fig:jedynka1}~b,c}) directions. \hyperref[fig:jedynka1]{Fig.~\ref*{fig:jedynka1}~a-c} present spectral intensity at Fermi energy ($E_F$) along semi-planar cuts of the reciprocal lattice. High symmetry points are given in~\hyperref[fig:jedynka1]{Fig.~\ref*{fig:jedynka1}~d} and experimental geometry is presented in~\hyperref[fig:jedynka1]{Fig.~\ref*{fig:jedynka1}~e}. Such an experiment allows to visualize not only~FS topography but enables a closer look into contributions from $f$~electrons and matrix element effects. The latter depend on experimental geometry, photon energy, probed Brillouin zone, etc. Angular windows of 36.5$^{\circ}$, 41$^{\circ}$ and 50$^{\circ}$ corresponding to variation along the $k'_x$ or $k_x$ axis in~\hyperref[fig:jedynka1]{Fig.~\ref*{fig:jedynka1}~a, b and c}, respectively, assured that different Brillouin zones were scanned with various experimental geometries. Hence, the differences between spectral intensities observed in different Brillouin zones originate from matrix element effects or a variation of $k_{\perp}$, which is a perpendicular to the surface component of the wave vector. The spectra recorded with $h\nu$=122~eV photons, namely at Ce~$4d$-$4f$ resonance, are characterized with an increased contribution from Ce~$4f$ electrons (\hyperref[fig:jedynka1]{Fig.~\ref*{fig:jedynka1}~a, b}) as compared to off-resonance data obtained with $h\nu$=88~eV~(\hyperref[fig:jedynka1]{Fig.~\ref*{fig:jedynka1}~c}). ARPES mapping with photon energy of 122~eV probes the region close to the $\Gamma$ point in the three-dimensional Brillouin zone~(\hyperref[fig:jedynka1]{Fig.~\ref*{fig:jedynka1}~d}) for normal emission~\citep{chen1}. The assumption of inner potential $V_0 = 16$~eV~\citep{chen1} allowed us to find the off-resonance energy of 88~eV, for which the scanned surface in k-space crosses the $\Gamma$ point  at $k_x$=0 and $k_y$=0 as well. 

\vspace{0.7cm}

Even a superficial analysis of spectral intensity dominated by Ce~$4f$ electrons (\hyperref[fig:jedynka1]{Fig.~\ref*{fig:jedynka1}~a, b}) leads to a conclusion that it breaks the reciprocal lattice symmetry and such a result is independent of the applied method of spectra normalization. Matrix element effects, which depend, among others, on experimental geometry (\hyperref[fig:jedynka1]{Fig.~\ref*{fig:jedynka1}~e}), are a reason for such an asymmetry. Moreover, they are responsible for different distributions of spectral intensity in \hyperref[fig:jedynka1]{Figs.~\ref*{fig:jedynka1}~a~and~b}, as these figures differ just by sample orientation. It is also known that a distribution of $f$-electron density and related spectral intensity depend on a form of hybridization between $f$~electrons and the valence band ($V_{cf}$)~\citep{ghaemi, starowicz} but such a distribution must fit into the reciprocal lattice symmetry in contrast to the matrix element effects. One can distinguish the areas of higher $4f$-electron signal, while other k-space regions exhibit a depletion of the intensity. Such high intensity areas are indicated with black dashed rectangles in~\hyperref[fig:jedynka1]{Fig.~\ref*{fig:jedynka1}~a-c}. It is noteworthy that the off-resonance FS~(\hyperref[fig:jedynka1]{Fig.~\ref*{fig:jedynka1}~c}) regains more reciprocal lattice symmetry in the region of normal emission ($k_x\sim 0$, $k_y\sim 0$), where fourfold symmetry is visible. 

The first approach to understand the spectral intensity are one-step model calculations, which realize a projection of both bulk and surface initial states on time-reversed low-energy electron diffraction final states~\citep{kkr}. The simulations were performed for three possible surface terminations (\hyperref[fig:jedynka1]{Fig.~\ref*{fig:jedynka1}~f}), namely, for Ce-In (\hyperref[fig:jedynka1]{Fig.~\ref*{fig:jedynka1}~g}), Co (\hyperref[fig:jedynka1]{Fig.~\ref*{fig:jedynka1}~h}) and In$_2$ (\hyperref[fig:jedynka1]{Fig.~\ref*{fig:jedynka1}~i}) surfaces as marked with arrows in~\hyperref[fig:jedynka1]{Fig.~\ref*{fig:jedynka1}~f}. The results indicate that Ce-In termination prevails as the corresponding simulated ARPES spectra for $h\nu$=122~eV (\hyperref[fig:jedynka1]{Fig.~\ref*{fig:jedynka1}~g}) fit the experimental results, what is also the case for the off-resonance ($h\nu$=88~eV) data (not shown). The simulations also exhibit a break of left-right symmetry, which is characteristic of an ARPES FS image. Thus, we observe that the matrix element effects are strong and they have to be separated if one wishes to single out intrinsic $c$-$f$ hybridization effects in a $4f$-electron spectrum.  

Experimental FS topography can be compared with KKR ground-state calculations yielding BSFs, which have been extracted in constant energy contours of $E_F$ ($E_B=0$) (\hyperref[fig:jedynka1]{Fig.~\ref*{fig:jedynka1}~j-o}) and 0.25~eV below $E_F$ (\hyperref[fig:jedynka1]{Fig.~\ref*{fig:jedynka1}~p-u}) for wave vector $k_z=0$ (\hyperref[fig:jedynka1]{Fig.~\ref*{fig:jedynka1}~j-u}). A presentation for two binding energies already gives some idea about dispersions, allows us to distinguish hole and electron bands, and helps us to find theoretical counterparts of the experimental FS. Total BSF distributions with suppressed $f$-electron contributions to valence bands are given in~\hyperref[fig:jedynka1]{Fig.~\ref*{fig:jedynka1}~j, p} and the corresponding plot for the system with itinerant $f$~electrons is shown in~\hyperref[fig:jedynka1]{Fig.~\ref*{fig:jedynka1}~k, q}. The calculations show a presence of hole pockets at $\Gamma$ and X and one large electron pocket at M. In fact, branches of experimental FS correspond to theoretical FS contours located either at $E_F$ or at $E_B=-0.25$~eV. The calculations for~$E_F$ show two hole pockets at~$\Gamma$, which are observed experimentally. The other parts of FS are well represented by the theoretical contours for $E_B=-0.25$~eV. This energy shift in the KKR calculations is well understood and is related to the limited angular momentum basis which leads to the uncertainty in the position of the Fermi level~\citep{kkr}. However, this does not affect the qualitative comparison between BSF and experimental data as well as the simulation of matrix element effects.

The difference between BSF for frozen (\hyperref[fig:jedynka1]{Fig.~\ref*{fig:jedynka1}~j, p}) and itinerant (\hyperref[fig:jedynka1]{Fig.~\ref*{fig:jedynka1}~k, q}) $f$ electrons shows an effect of $c$-$f$ hybridization on electronic structure. The most prominent changes are located around the ${\Gamma}$ point and along the $\Gamma$-$X$ direction. To visualize a role of particular atoms and their contribution to the FS, BSF for itinerant electrons have been projected on atomic wave functions~(\hyperref[fig:jedynka1]{Fig.~\ref*{fig:jedynka1}~l-o, r-u}). Ce atoms donate mainly $4f$~electrons at~$E_F$ with high contribution in the region of $\Gamma$ and X points, where large FS modifications appear with the introduction of $f$~electrons to the valence band. This area is quite affected by $c$-$f$ hybridization effects. It is visible, in particular at~$E_F$~(\hyperref[fig:jedynka1]{Fig.~\ref*{fig:jedynka1}~k,l}). Such $f$-electron spectral intensity is diminished below $E_F$~(\hyperref[fig:jedynka1]{Fig.~\ref*{fig:jedynka1}~q,r}). Electrons from Co (mainly $3d$) and In atoms from In planes shown by black arrows in~\hyperref[fig:jedynka1]{Fig.~\ref*{fig:jedynka1}~f} (high contribution from $5p$) contribute, rather, to the whole FS. However, Co $3d$ states have a high population in the electron pockets around M (well seen for $E_B=-0.25$~eV), while this population is reduced in some bands around $\Gamma$. Interestingly, In $5p$ electrons from Ce-In planes (green arrows in~\hyperref[fig:jedynka1]{Fig.~\ref*{fig:jedynka1}~f}) exhibit four prominent maxima around the $\Gamma$ point visible, in particular, at $E_B=-0.25$~eV. These maxima correspond to the location of four hot spots in ARPES maps located around the $\Gamma$ point along the $\Gamma$-M direction~(\hyperref[fig:jedynka1]{Fig.~\ref*{fig:jedynka1}~a-c}). BSFs represent states from the whole Brillouin zone, whereas one-step model calculations show that the contribution from Ce-In surface dominates the ARPES image. Even if four In atoms per unit cell give high intensity in BSF, one indium atom per unit cell located in the surface Ce-In planes may deliver quite significant contribution to ARPES spectra. The hot spots observed in ARPES are found just at the places with increased contribution from In located in Ce-In planes. Moreover, one can see that also Ce~$4f$ electrons yield a high amplitude of BSF at the same points in k-space (\hyperref[fig:jedynka1]{Fig.~\ref*{fig:jedynka1}~r}). This will be reflected in high $c$-$f$ hybridization effects observed by ARPES. The experimental FS obtained in this paper reproduces some shapes also present in previous ARPES studies~\citep{koicz,dudy,chen1,jang,reber}. However, FSs obtained before were recorded at higher temperatures and, depending on a report, with other photon energies, experimental geometry and in some cases other surface terminations. Hence, also four hot spots around $\Gamma$ were not discussed before.

\vspace{0.3cm}
\begin{figure*}
\includegraphics[width=0.95\textwidth]{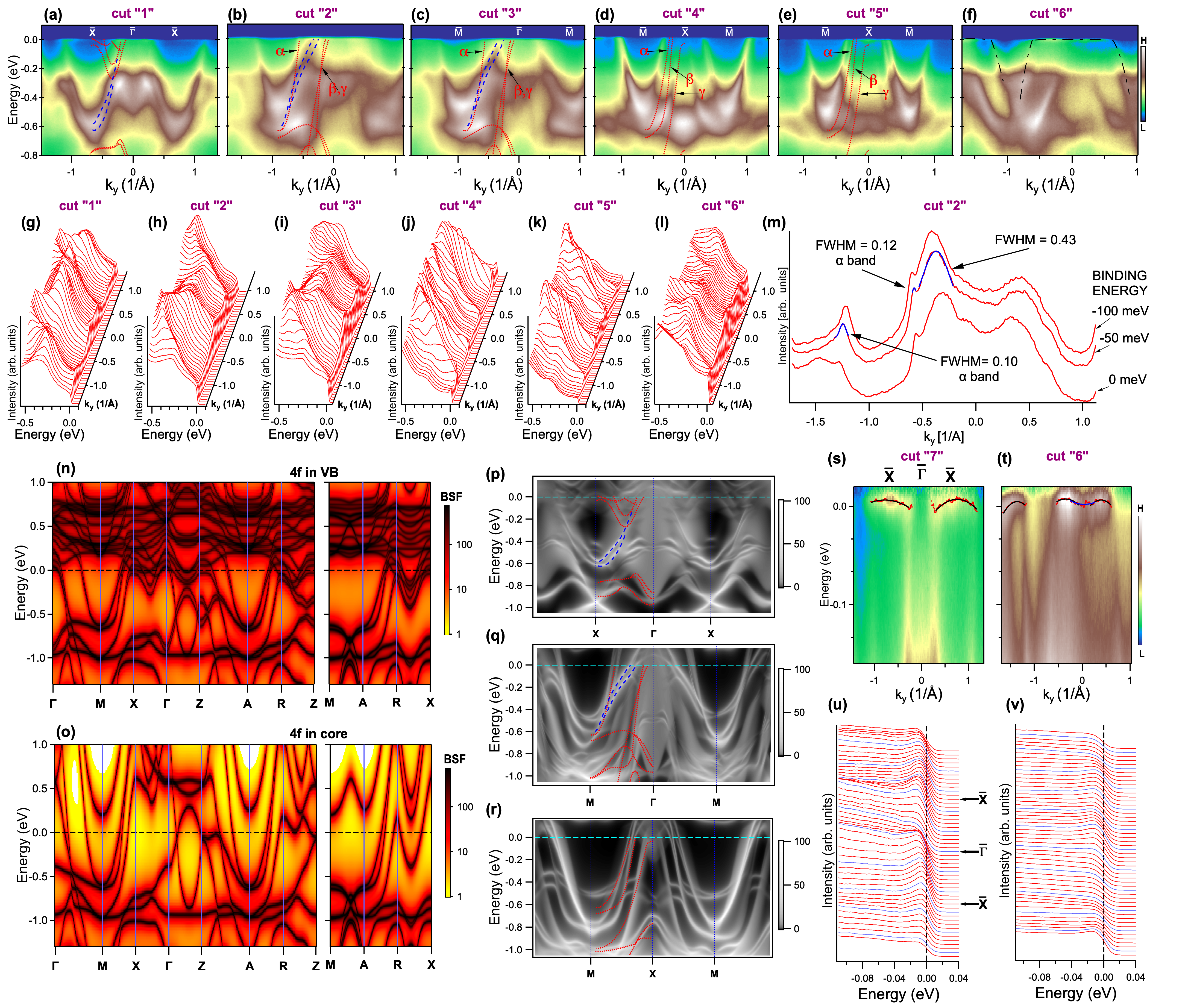}
\caption{Band structure of CeCoIn$_5$ along selected paths in k-space denoted as cut 1 -- cut 7 in \hyperref[fig:jedynka1]{Fig.~\ref*{fig:jedynka1}}. ARPES scans recorded with photon energy of 122 eV at $T=6$~K are shown along the paths: 1 (\={X}-$\bar{\Gamma}$-\={X}) (a), 2 (\={M}-$\bar{\Gamma}$-\={M}) (b),~3  (c),~4 (\={M}-\={X}-\={M}) (d), ~5  (\={M}-\={X}-\={M}) (e), and  6 (f). The same band structures are represented in a form of EDCs in the corresponding panels: (g) - (l). Red dotted lines represent bulk CeCoIn$_5$ dispersions obtained from BSF fitting, whereas blue dashed lines denote surface states, which appeared in one-step model simulations. The black dash-dot line in (f) is a guide to the eye, which shows the dispersion. Selected MDCs extracted from (b) are shown in (m) with full width at half maximum (FWHM) of the fitted Gaussian peaks. Panels (n) and (o) present BSFs along important crystallographic directions calculated with SPR-KKR package in presence ($4f$ in VB) and absence ($4f$ in core) of $c$-$f$ hybridization effects, respectively. One-step model of photoemission simulations of ARPES spectra are presented for photon energy of 122 eV and Ce-In surface termination along X-$\Gamma$-X (p), M-$\Gamma$-M (q) and M-X-M (r) directions. The meaning and positions of red dotted and blue dashed lines is the same as in (a)-(e). Weakly dispersive band is visible near the Fermi level in spectra collected for cuts 7 (s) and 6 (t) shown in \hyperref[fig:jedynka1]{Fig.~\ref*{fig:jedynka1}} after dividing by Fermi-Dirac distribution (FDD). Fitted dispersions are represented by black lines. The corresponding EDCs shown in (u) and (v) have not been normalized with FDD.}
\label{fig:dwojka1}
\end{figure*}
\vspace{0.5cm}

Band structure obtained by ARPES is shown along important crystallographic directions, as well as along other selected paths in k-space (\hyperref[fig:dwojka1]{Fig.~\ref*{fig:dwojka1}~a-f}). These paths are numbered and drawn on experimental FSs (\hyperref[fig:jedynka1]{Fig.~\ref*{fig:jedynka1}~a,b}). The increased $f$-electron contribution reflected in a high signal at~$E_F$ is well visible in the same spectra presented as energy distribution curves (EDCs) in the corresponding panels: (\hyperref[fig:dwojka1]{Fig.~\ref*{fig:dwojka1}~g-l}). Momentum distribution curves (MDCs) for the cut 2 visualize that the peak width varies considerably depending on a band (\hyperref[fig:dwojka1]{Fig.~\ref*{fig:dwojka1}~m}). One can resolve an electron pocket around \={M} (\hyperref[fig:dwojka1]{Fig.~\ref*{fig:dwojka1}~b-e}), which is named $\alpha$~\citep{dudy,chen1,jang}. We also observe hole pockets around the $\bar{\Gamma}$ point (\hyperref[fig:dwojka1]{Fig.~\ref*{fig:dwojka1}~a-c}) and two holelike dispersions observed at the \={X} point along the \={M}-\={X}-\={M} direction (\hyperref[fig:dwojka1]{Fig.~\ref*{fig:dwojka1}~d, e}). These dispersions have been attributed in previous studies to $\beta$ and $\gamma$ bands~\citep{dudy,chen1,jang} but this assignment is modified in the current report. $\alpha$, $\beta$ and $\gamma$ bands were named 135, 133 and 131, respectively in previous studies~\citep{koicz}. 

To interpret the spectra we present calculated BSFs (\hyperref[fig:dwojka1]{Fig.~\ref*{fig:dwojka1}~n}) with $f$~electrons in valence band. To simulate the effect of $f$~electrons on band structure, we have also calculated BSFs with frozen $f$~electrons (\hyperref[fig:dwojka1]{Fig.~\ref*{fig:dwojka1}~o}). It is visible that when $f$~electrons are added to the valence band, heavy-particle bands appear in the vicinity of~$E_F$. Detailed projections of BSF on particular atoms along high symmetry directions are shown in the~\hyperref[app:b]{Appendix~\ref*{app:b}}. The obtained BSFs (\hyperref[fig:dwojka1]{Fig.~\ref*{fig:dwojka1}~n}) are in qualitative agreement with other calculations performed by the VASP package using strongly constrained and appropriately normalized semilocal density functional (not shown)~\citep{vasp,scan}. 

It is apparent that the measured band structure along the \={X}-$\bar{\Gamma}$-\={X} direction is not reproduced by calculated BSF, namely, the experimental electronlike band around the \={X} point is missing in theoretical calculations. Moreover, similar spectra along \={X}-$\bar{\Gamma}$-\={X} have been obtained before and not reproduced by DFT calculations performed for bulk CeCoIn$_5$~\cite{chen1,reber}. To clarify the origin of this band and search for possible contribution of surface states, we performed one-step model calculations for Ce-In surface termination along X-$\Gamma$-X (\hyperref[fig:dwojka1]{Fig.~\ref*{fig:dwojka1}~p}), M-$\Gamma$-M (\hyperref[fig:dwojka1]{Fig.~\ref*{fig:dwojka1}~q}), and M-X-M (\hyperref[fig:dwojka1]{Fig.~\ref*{fig:dwojka1}~r}) directions. 

Both BSF and one-step model calculations have been realized for the same potential distribution. It is instructive to compare these results as new dispersions, which appear in one-step model calculations, can be related to surface states. Thus, the analysis of \hyperref[fig:dwojka1]{Fig.~\ref*{fig:dwojka1}~p-r} is a clue for the interpretation of the ARPES spectra. The bulk dispersions extracted from BSF (\hyperref[fig:dwojka1]{Fig.~\ref*{fig:dwojka1}~n}) are shown as red dotted lines in \hyperref[fig:dwojka1]{Fig.~\ref*{fig:dwojka1}~p-r}. It is obvious that bulk features do not cover all dispersions. In particular, new dispersions yielding high intensity are found along X-$\Gamma$ (\hyperref[fig:dwojka1]{Fig.~\ref*{fig:dwojka1}~p}) and M-$\Gamma$ (\hyperref[fig:dwojka1]{Fig.~\ref*{fig:dwojka1}~q}) directions and are marked with blue dashed lines. These are interpreted as additional dispersions related to surface states for Ce-In surface termination. The same curves representing theoretical bulk (red dotted lines) and surface (blue dashed lines) state dispersions are superimposed on ARPES data (\hyperref[fig:dwojka1]{Fig.~\ref*{fig:dwojka1}~a-e}), indicating the origin of measured dispersions. 

Thus, it is found that paraboliclike band around the \={X} point in \hyperref[fig:dwojka1]{Fig.~\ref*{fig:dwojka1}~a} comes from the surface state. Moreover, it appears that high intensity hole pocket around $\bar{\Gamma}$ is located at dashed blue lines (cf. \hyperref[fig:dwojka1]{Fig.~\ref*{fig:dwojka1}~b,c}), then it should originate or at least have a large contribution from surface states related to Ce-In termination. It seems that along \={M}-\={X}-\={M} (\hyperref[fig:dwojka1]{Fig.~\ref*{fig:dwojka1}~d,e}) we do not observe a substantial contribution from surface states near $E_F$. One of the characteristic features of surface states is their two-dimensional character. Such a character can be verified by scans with different photon energies to test eventual dispersion in the direction perpendicular to surface. Such analysis is shown in~\hyperref[app:a]{Appendix~\ref*{app:a}}. It is found that the peak formed by $\beta$ and $\gamma$ and surface states along \={M}-$\bar{\Gamma}$-\={M} is more complex and may have a high intensity nondispersing component indeed. 

Previous ARPES measurements performed by Chen et al.~\citep{chen1} delivered interesting information on heavy fermion formation in CeCoIn$_5$. Namely, these studies have shown that along the M-$\Gamma$ direction the bands $\beta$ and $\gamma$ (as named in Ref.~\citep{chen1}) become broad and gain a high $f$-electron related intensity with lowering temperature down to 17~K. We had the opportunity to measure the same region at a much lower temperature of 6~K. Therefore, the hybridization effects are more pronounced. However, we find that the bands, which are much broadened and yield high intensity hot spots at the FS, are related to Ce-In surface or at least have a dominating surface state contribution. 

On the other hand, $\alpha$ band, which is well resolved, remains sharp and represents weakly correlated electrons. This is visible in the analysis of exemplary MDCs (\hyperref[fig:dwojka1]{Fig.~\ref*{fig:dwojka1}~m}) obtained in a single measurement assuring the same experimental conditions for the band structure along cut 2 located next to the \={M}-$\bar{\Gamma}$-\={M} direction (\hyperref[fig:jedynka1]{Fig.~\ref*{fig:jedynka1}}). A similar MDCs image is found for the \={M}-$\bar{\Gamma}$-\={M} direction (not shown). It is found that high intensity dispersion with surface state has a large peak width (0.4 \AA$^{-1}$) in contrary to the $\alpha$ band (0.1 \AA$^{-1}$). Due to such broadening two surface dispersions, obtained in theory, are not resolved. 

\hyperref[fig:dwojka1]{Fig.~\ref*{fig:dwojka1}~d,~e} present scans along \={M}-\={X}-\={M} measured in different Brillouin zones, which differ by a certain value of $k_z$ vector. $\alpha$ and $\beta$ bands are better separated in~\hyperref[fig:dwojka1]{Fig.~\ref*{fig:dwojka1}~e}. Spectral intensity in these panels also differs due to matrix element effects involving $4f$ electrons, namely, larger contribution from Ce~$4f$ electrons is visible below $k_y\sim-1$ \AA$^{-1}$ and above $k_y\sim~1$~\AA~$^{-1}$  in~\hyperref[fig:dwojka1]{Fig.~\ref*{fig:dwojka1}~d,j}, while it is enhanced around X in~\hyperref[fig:dwojka1]{Fig.~\ref*{fig:dwojka1}~e,k}. This agrees with the general tendency of high and low intensity of $4f$ electrons related to matrix element effects, what is observed in \hyperref[fig:jedynka1]{Fig.~\ref*{fig:jedynka1}~a-c}, where high $f$-electron intensity regions are marked with rectangles. Moreover, typical effects of $c$-$f$ hybridization are also visible in many spectra and the example is shown in~\hyperref[fig:dwojka1]{Fig.~\ref*{fig:dwojka1}~f}. This image presents the dispersion (black dash-dotted line) and spectral shape characteristic of $c$-$f$~hybridization effect described by the periodic Anderson model~\citep{tahvildar}, namely, the band approaches $E_F$, where it gains more $f$-electron character and flat dispersion.

Finally, a dispersion of $f$-electron bands has been found. The spectra normalized by Fermi-Dirac distribution (FDD) were subjected to EDC fitting~(\hyperref[fig:dwojka1]{Fig.~\ref*{fig:dwojka1}~s,t}). The extracted dispersions were approximated by parabolas, which allowed us to estimate effective masses. The dispersions in \hyperref[fig:dwojka1]{Fig.~\ref*{fig:dwojka1}~o} yield the values from 70 to 130 free electron masses ($m_e$), whereas the fits shown in \hyperref[fig:dwojka1]{Fig.~\ref*{fig:dwojka1}~t} give 30 to 40 $m_e$ for holelike and 80~$m_e$ for electronlike dispersions. This does not exclude the existence of higher or lower effective masses in other regions of the Brillouine zone (BZ). The obtained effective mass of the order of ~$10^2$ $m_e$ would agree with the Sommerfeld coefficient estimated for CeCoIn$_5$ of~$\gamma =$ 290~mJ/(mole$\times$K$^2$)~\citep{petrovic}. Importantly, heavy electron dispersions are also visible in raw data, which are not normalized to FDD~(\hyperref[fig:dwojka1]{Fig.~\ref*{fig:dwojka1}~u,v}).

\begin{figure*}
\centering\includegraphics[width=0.95\textwidth]{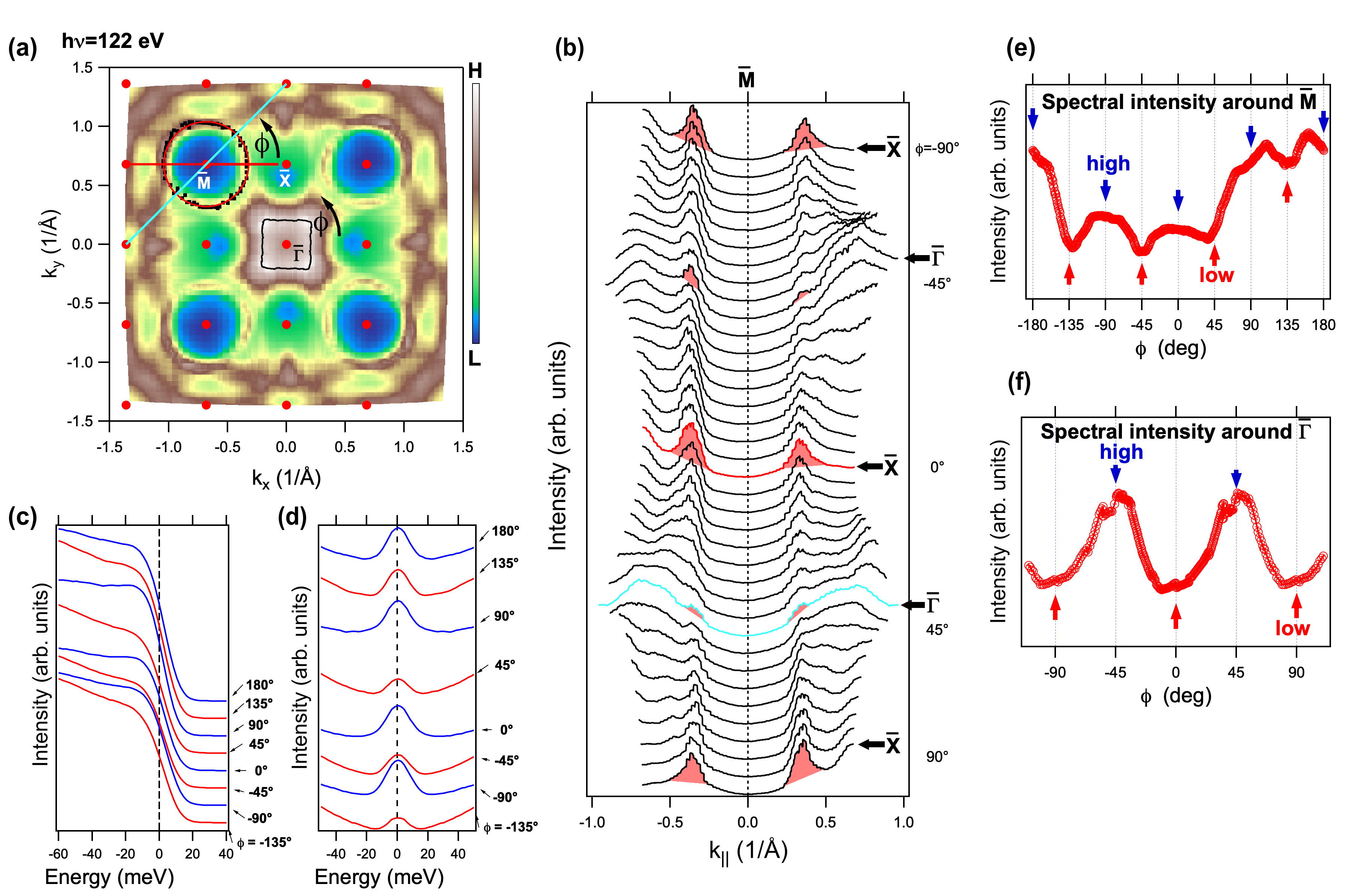}
\caption{(a) Symmetrized experimental Fermi surface obtained by a superposition of ARPES data rotated 4 times by 90$^{\circ}$. Red circle presents the pocket around \={M} subjected to $f$-electron intensity analysis. $\phi$ is defined as the angle between horizontal \={M}-\={X} direction and the line crossing the particular point of the contour and \={M} point, measured in counterclockwise direction. (b) Momentum distribution curves extracted from symmetrized data along lines crossing \={M} point in different directions given by $\phi$. The exemplary red and cyan curves represent paths which are drawn with the same colors in (a). The peaks related to the contour around \={M} are highlighted with red area for \={M}-\={X} and $\bar{\Gamma}$-\={X} directions. (c) Energy distribution curves (EDCs) and (d) symmetrized EDCs from the electron pocket around \={M} for selected $\phi$ angles. (e) Intensity at the Fermi energy along the electron pocket indicated with the red circle in (a) as a function of $\phi$. Red arrows in~(e) denote the minima of intensity, which appear to be located along $\bar{\Gamma}$-\={M} direction, while blue arrows show \={M}-\={X} direction characterized with higher spectral intensity. (f) Intensity at the Fermi energy along the hole pocket around $\bar{\Gamma}$ indicated by the black contour in (a) as a function of $\phi_1$. Red arrows denote minima in intensity, which are complementary to those in (e).}
\label{fig:trojka1}
\end{figure*}

Anisotropic hybridization should affect ARPES spectra considerably. Varying contributions of $4f$~electrons to the spectra along the FS should be the primary effects. Qualitatively, an admixture of the conduction Bloch state to the $f$-electron states expresses the hybridization magnitude $V_{cf}$. Thus, $c$-$f$ hybridization directly influences spectral intensity at~$E_F$ together with matrix element effects. The latter should be separated out to have a bare contribution of $4f$~electrons to the spectra related to the $c$-$f$~hybridization. To remove partially geometrical effect of matrix elements, the FS was symmetrized by adding spectra rotated by 90$^{\circ}$, 180$^{\circ}$ and 270$^{\circ}$ to the original FS~(\hyperref[fig:trojka1]{Fig.~\ref*{fig:trojka1}~a}). We have also performed an alternative symmetrization procedure using reflections or their combination with rotations (what corresponds to full 4/$mmm$ point symmetry of $\Gamma$ point), but this resulted in an image of a worse quality. The contour of the FS corresponding to an electron pocket around the \={M} point has been described by the circle which was fitted to points belonging to the FS~(\hyperref[fig:trojka1]{Fig.~\ref*{fig:trojka1}~a}). The MDCs crossing the M point and corresponding to different $\phi$ angles have been extracted at the Fermi energy~(\hyperref[fig:trojka1]{Fig.~\ref*{fig:trojka1}~b}). The integration over the energy window of $\pm$20~meV around~$E_F$ was applied. A closer inspection shows that higher intensity of $f$-electron related peak at $E_F$ is present along the \={M}-\={X} direction, whereas it is lower at \={M}-$\bar{\Gamma}$. The EDCs shown for the contour around the \={M} point exhibit a variation indicating higher $f$-electron contribution (for blue curves) along the same directions as for MDCs~(\hyperref[fig:trojka1]{Fig.~\ref*{fig:trojka1}~c}). This is visible in higher intensity close to the Fermi level at the binding energy where typically the $f$-electron spectrum is located. The increased contribution near $E_F$ is also typically seen in the symmetrized blue EDCs~(\hyperref[fig:trojka1]{Fig.~\ref*{fig:trojka1}~d}). Finally, the integrated (over 0.1~x~0.1~\AA$^{-2}$ square surroundings) intensity along the electron pocket roughly reproduces minima along \={M}-$\bar{\Gamma}$ and maxima along \={M}-\={X} directions~(\hyperref[fig:trojka1]{Fig.~\ref*{fig:trojka1}~e}). Hence, the symmetrized experimental FS yields the variation of spectral intensity around the \={M} point with partially reduced matrix element effects. Moreover, \hyperref[fig:trojka1]{Fig.~\ref*{fig:trojka1}~a} presents a variation of spectral intensity around the $\bar{\Gamma}$ point. This shows maxima along $\Gamma$-\={M} and minima along $\bar{\Gamma}$-\={X} for $4f$ electron contribution. The spectral intensity at~$E_F$ (\hyperref[fig:trojka1]{Fig.~\ref*{fig:trojka1}~f}) is plotted along the rectangular contour (\hyperref[fig:trojka1]{Fig.~\ref*{fig:trojka1}~a}) around the $\bar{\Gamma}$ point. We did not analyze MDCs and EDCs in this case, as more bands are present around $\bar{\Gamma}$. For the on-resonant Ce $4d$-$4f$ photoemission the spectra are dominated at $E_F$ by $4f$ electrons. Here it may be considered that such variations of spectral intensity observed around \={M} and $\bar{\Gamma}$ points may reflect a real variation of $f$-electron density in momentum space. 

The hypothesis that the distribution of $f$-electron density has been found can be verified by tight-binding (TB) model calculations. One should keep in mind that ARPES spectra have a considerable contribution from surface termination, which are Ce-In planes as was proved by comparison with one-step model calculations~(\hyperref[fig:jedynka1]{Fig.~\ref*{fig:jedynka1}}~and~\hyperref[fig:dwojka1]{Fig.~\ref*{fig:dwojka1}}). Moreover, the highest contribution to the FS from In located in Ce-In planes ($5p$ states) is found at 4 hot spots around the $\bar{\Gamma}$ point, where strong hybridization, leading to high spectral broadening at low temperature, is observed. These arguments support the assumption to consider a simple model of Ce-In planes in the TBA calculations. Here we neglect the out of plane hybridization, which will not have a strong impact on spectral intensity distribution along the plane. Also the hybridization with Ce $5d$ orbitals is not taken into account. 

The idea of a minimal TBA model for CeCoIn$_5$ appeared earlier~\citep{maehira}. However, previously, the model has been fitted to DFT calculations. In this paper, the TBA parameters have been found by fitting theoretical dispersions to the experimental binding energies at \={M} and \={X} points, as well as to experimental Fermi wave vectors in the $\bar{\Gamma}$-\={X} and \={X}-\={M} directions. The obtained theoretical bands are shown with measured ARPES spectra along \={M}-\={X}-\={M} and \={X}-$\bar{\Gamma}$-\={X} directions (~\hyperref[fig:czworka]{Fig.~\ref*{fig:czworka}~a-b}), which correspond to the paths 5 and 1 in \hyperref[fig:jedynka1]{Fig.~\ref*{fig:jedynka1}}, respectively. The size of the marker is proportional to the $4f$-electron contribution to the band. One can see that TBA model describes well the electron pocket around the \={M} point. Several bands with dominating $f$-electron contribution are visible above the Fermi level. It is noteworthy that calculations predict the presence of a hole band around $\Gamma$ point~(\hyperref[fig:czworka]{Fig.~\ref*{fig:czworka}~b}), which becomes flat and gains $f$-electron character close to the Fermi level. This is obvious that not every band visible in the measured spectra can be identified in the theory as the TBA minimal theoretical model has to reproduce $f$-electron intensity variation originating from surface Ce-In layer. 

A particular attention should be paid to the distribution of the spectral weight around $\bar{\Gamma}$ point, which, in accordance with ours and previous studies, is related to the heavy quasiparticle band developing at low temperature. The distribution of spectral weight at FS predicted by the model is shown in~\hyperref[fig:czworka]{Fig.~\ref*{fig:czworka}~c}. The projections on different base states have been shown in~\hyperref[fig:czworka]{Fig.~\ref*{fig:czworka}~d-h}. Both almost rectangular hole pocket around the $\bar{\Gamma}$ point and electron pocket around M exhibit specific variation of $f$-electron spectral intensity at $E_F$. Such spectral intensity was summed over all contributing $f$ orbitals in~\hyperref[fig:czworka]{Fig.~\ref*{fig:czworka}~(i,j)}. $f$-electron intensity around $\bar{\Gamma}$ reaches its maximum at the $\bar{\Gamma}$-\={M} direction, while the minima appear at the \={X}-$\bar{\Gamma}$ direction. The spectral weight visible in the contour around M point (corresponding to the electron pocket) shows the complementary pattern. On the other hand, the distribution corresponding to In states is much more uniform at the FS. It should be remarked that the highest contribution to the FS comes from $f_a$ and $f_c$ orbitals, corresponding to $J_z$ values of $\pm \frac{5}{2}$ and  $\pm \frac{3}{2}$, respectively.
These orbitals determine the ground-state symmetry of bulk CeCoIn${_5}$ according to previous studies realized by means of linearly polarized x-ray absorption spectroscopy and the nonresonant inelastic x-ray scattering~\citep{willers2015,sundermann2019}.

The observed variation of the spectral weight is consistent with the results of our experiment. It is noteworthy that the maxima and minima of $4f$ related experimental intensity shown in~\hyperref[fig:trojka1]{Fig.~\ref*{fig:trojka1}} are reproduced by TBA calculations both around $\bar{\Gamma}$ and \={M} points. Moreover, a series of fits for different TBA parameters, including those from Maehira et al.~\citep{maehira} (not shown), revealed the same tendency (see also~~\hyperref[app:c]{Appendix~\ref*{app:c}}). The agreement between the experiment and theory is in favor of the thesis that $f$-electron density in momentum space has been found, where quite large spectral contribution comes from surface Ce-In planes. However, one should be careful with such an approach, as spectral intensity in ARPES is quite complex and the used symmetrization does not eliminate all matrix element and final state effects. Also, a quite important contribution of non $f$-electrons remains in the spectra even for on-resonance studies.

\begin{figurehere}
\centering\includegraphics[width=0.95\linewidth]{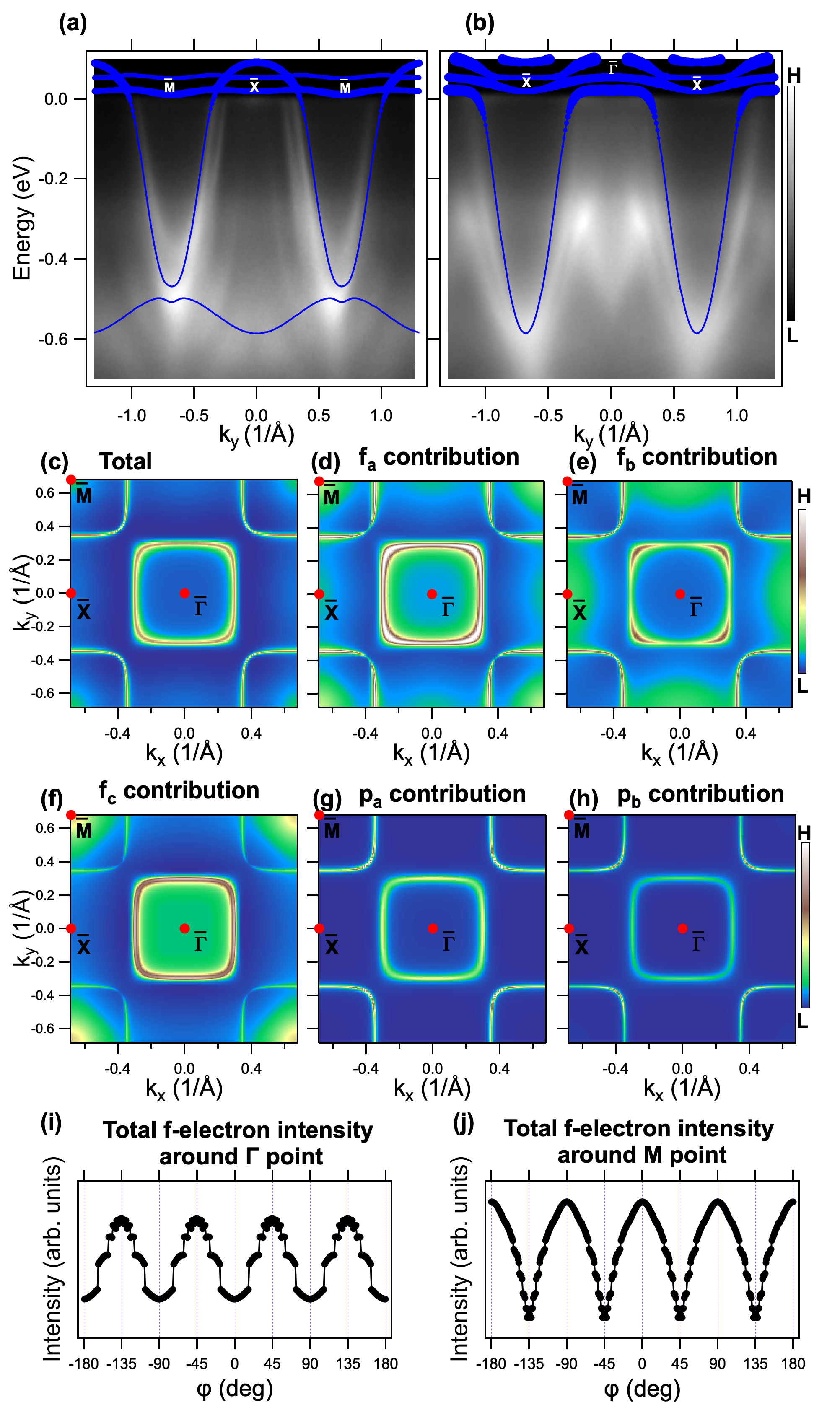}
\caption{Theoretical band structure for Ce-In planes (solid lines) obtained by tight-binding (TBA) calculations. TBA parameters have been adjusted by fitting theoretical bands (solid blue lines) to ARPES data collected along path~5 and~1, what corresponds to \={M}-\={X}-\={M} (a) and \={X}-$\bar{\Gamma}$-\={X} direction, respectively. The marker size is proportional to the contribution of $f$ orbital to the band. Panel (c) shows the distribution of the spectral weight on the Fermi surface. The projections on different base states are shown in (d)-(h). The Lorentzian broadening of 15~meV was applied in (c)-(h) maps. Panels (i) and (j) show total $f$-electron intensity extracted from TBA calculations along FS contours around $\Gamma$ and M points. Azimuthal angle $\phi$=0 corresponds to $\bar{\Gamma}$-\={X} direction in (i) and \={M}-\={X} direction in (j).}
\label{fig:czworka}
\end{figurehere}

\section{Conclusions}
The electronic structure of CeCoIn$_5$ system has been studied with resonant ARPES at low temperature, using the Ce~$4d$-$4f$ transition. The determined topography of band structure is in a general agreement with that reported previously~\citep{koicz,dudy,chen1,jang}. Relativistic multiple scattering KKR and the one-step model calculations were able to simulate the ARPES data and the best compatibility was reached for the assumed Ce-In surface termination. These calculations allowed us to single out surface states, which contribute with high intensity to ARPES spectra. We have observed various effects of $c$-$f$ hybridization between $f$~electrons and valence band electrons. These effects include a presence of heavy effective mass states near the Fermi energy and bands gaining strong $f$-electron character near $E_F$. The spectra also reveal a coexistence of bands with narrow and broadened MDCs and these latter were attributed to surface states. Comparison with theoretical calculations shows that the most significant hybridization effects present in the spectra are related to Ce-In planes and such statement concerns ARPES data interpretation and not bulk properties. We observe a considerable variation of $f$-electron spectral intensity in momentum space, originating from both, matrix element effects and momentum-dependent hybridization. A symmetrization of FS obtained experimentally in a wide angular range allows us to reduce the influence of matrix element effects and yields spectral intensity with high $f$-electron contribution. The same structure of momentum dependence of $f$-electron density is the outcome of the TB model calculations realized for Ce-In plane for a variety of parameters. Thus, the theoretical modeling is in favor of the hypothesis that variation of $f$-electron density in momentum space dominates in the symmetrized spectra and Ce-In surface layer yields substantial contribution to the spectra.

\begin{acknowledgments}
This work has been supported by~~the~~National~~Science Centre, Poland within Grants No. 2016/23/N/ST3/02012 and~No.~2015/19/B/ST3/03158. J.S., M.F., and R.K. acknowledge the financial  support from Grant OPUS, No. UMO-2018/29/B/ST3/02646 from the National~Science~Centre,~Poland. Support of the Polish Ministry of Science and Higher Education under the grant 7150/E-338/M/2018 is acknowledged.  J.M. and L.N. would like to thank the CEDAMNF (CZ.02.1.01/0.0/0.0/15\_003/0000358) cofunded by the Ministry of Education, Youth and Sports of Czech Republic. This work is funded by the Deutsche Forschungsgemeinschaft (DFG, German Research Foundation) through Project-ID 258499086 - SFB 1170 (C06). We acknowledge financial support from the DFG through the W\"urzburg-Dresden Cluster of Excellence on Complexity and Topology in Quantum Matter -- \textit{ct.qmat} (EXC 2147, Project-ID No. 39085490). D.L and A.P.K. acknowledge support of the European Regional Development  Fund in the IT4Innovations National Supercomputing Center - Path to Exascale Project, No. CZ.02.1.01/0.0/0.0/16\_013/0001791 within the Operational Programme Research, Development and Education and the project e-INFRA CZ (ID:90140) by the Ministry of Education, Youth and Sports of the Czech Republic. D. L. also acknowledges Project No. EHP-CZ-ICP-1-013.
\end{acknowledgments}

\bibliographystyle{IEEEtran}
\bibliography{mybibfile115}

\appendix
\section{Photon energy dependence of ARPES spectra}
\label{app:a}

ARPES studies were performed for different photon energies in the range between h$\nu$=122 eV and 105 eV, which correspond to $\Gamma$ and Z points at normal emission, respectively. This is estimated taking into account the inner potential of $V_0 = 16$~eV~\citep{chen1}. MDCs obtained along \={M}-$\bar{\Gamma}$-\={M} direction at $E_F$ are shown in \hyperref[fig:photon]{Fig.~\ref*{fig:photon}}. The scan obtained with h$\nu$=122 eV was realized at k$'_y=0$ line in ~\hyperref[fig:jedynka1]{Fig.~\ref*{fig:jedynka1}~a}. Such measurements can visualize eventual dispersions along the perpendicular to the surface component of wave vector $k_z$, which changes with photon energy. MDCs have been analyzed by fitting three peaks on both sides of $\bar{\Gamma}$. Unfortunately the peaks are not well resolved in the spectra. The highest intensity peak is shown with filled circles. It corresponds to $\beta$, $\gamma$ bands and surface states but the last should dominate. $\alpha$ band is located on both sides of \={M} and shown with crosses. These bands were introduced in \hyperref[fig:dwojka1]{Fig.~\ref*{fig:dwojka1}~b}. The third, small intensity fitted peak is located closer to $\bar{\Gamma}$ point (not shown). We do not see clear dispersion of the highest intensity peak, which is convergent with the hypothesis that it originates mainly from surface states

\begin{figurehere}
\centering\includegraphics[width=0.95\linewidth]{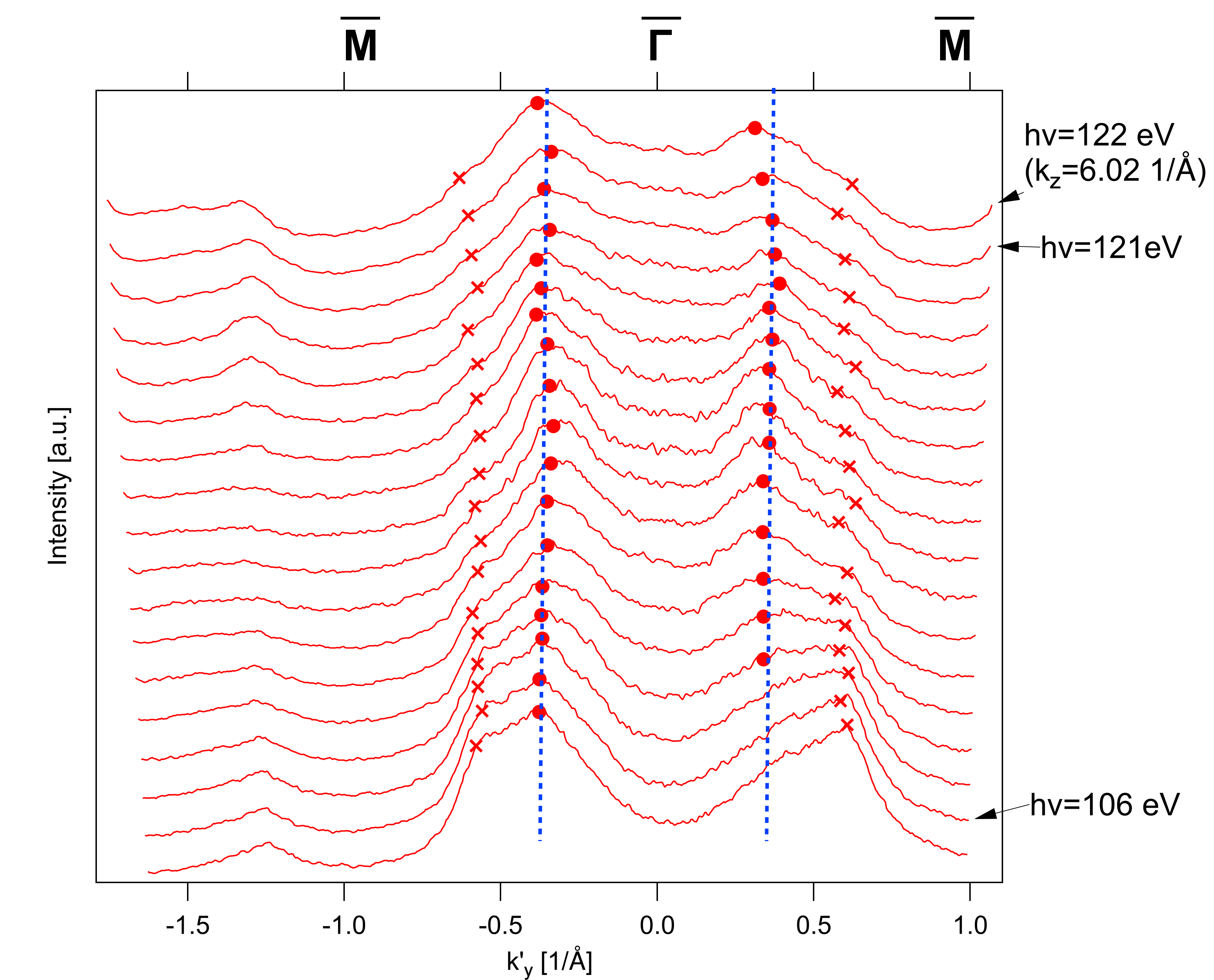}
\caption{MDCs at $E_F$ obtained with ARPES along \={M}-$\bar{\Gamma}$-\={M} direction for CeCoIn$_5$. The measurements were performed at $T=6$~K as a function of photon energy in the range of 105 eV to 122 eV with a step of 1 eV. Such energy range corresponds to a variation of perpendicular to surface component of wave vector $k_z$ between 5.63 and 6.02~\AA$^{-1}$. Circles and crosses represent positions of the fitted peaks. Blue dashed line is drawn to visualize possible lack of dispersion of the higher intensity peak. Crosses indicate the position of $\alpha$ band.} 
\label{fig:photon}
\end{figurehere} 

\newpage
\section{Band structure of CeCoIn$_5$ calculated along high symmetry directions using DFT method}
\label{app:b}
\begin{figure*}
\centering\includegraphics[width=0.8\linewidth]{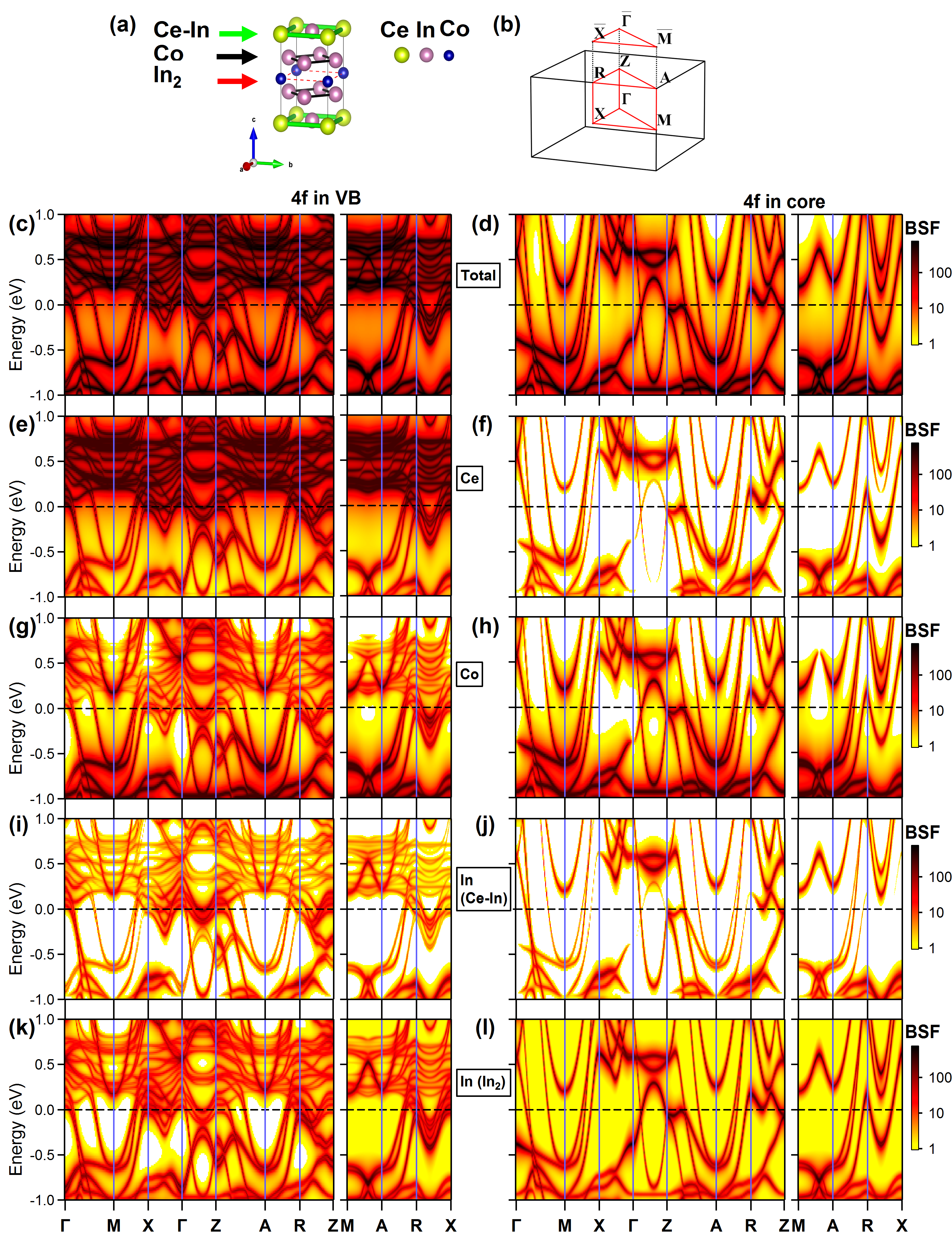}
\caption{Bloch spectral functions calculated along high symmetry directions in the BZ.  (a) The crystal structure of CeCoIn$_5$. (b) The first Brillouin zone with marked high symmetry points. Two types of calculations are provided: with included hybridization between $4f$ states and rest of orbitals (c,e,g,i,k), as well as the results without the hybridization ($4f$ in core) (d,f,h,j,l). The projections on Ce (e,f), Co (g,h), In from Ce-In planes (i,j) and In from In planes (k,l) are also provided.} 
\label{fig:bsf}
\end{figure*}

Calculated BSFs for CeCoIn$_5$ along high symmetry directions are shown in~\hyperref[fig:bsf]{Fig.~\ref*{fig:bsf}~c-l}.  Projections on different atoms in unit cell have been calculated. To analyze the effect of $f$~electrons on matrix elements, we performed several calculations with limited angular momentum expansion of the initial state Green's function ($l_{max}=2$ which includes $s$, $p$ and $d$ orbitals). These calculations are shown in~\hyperref[fig:jedynka1]{Fig.~\ref*{fig:jedynka1}~j-u} of the article and commented  either with $4f$ electrons in valence band ("$4f$ in VB") or with $4f$ electrons excluded from the valence band ("$4f$ in core"). The results obtained with the second approach can be interpreted as the electronic structure in the absence of hybridization effects. Therefore, the differences between results obtained in these two ways are directly  related to the $V_{cf}$ hybridization effects. Please note that in the case of CeCoIn$_5$, hybridization effects between $f$ and $d$ states can be very well described by the single particle approach. Additional correlation effects beyond LDA approach, which could be included by the LDA+U or LDA+DMFT methods, do not change the conclusion concerning matrix element effects discussed in the article. 
  
\newpage

\section{Tight-binding approach with the Slater-Koster analysis}
\label{app:c}
\subsection{Tight-binding model}

The computed DFT intensity profiles of the ARPES spectra for $\mathrm{CeCoIn_5}$ are sensitive to the details of the system surface, with the best overall agreement for $\mathrm{Ce}$-$\mathrm{In}$ termination. A complementary insight into the structure and symmetry of hybridization between Ce 4$f$ and the $5p$ conduction electrons due to indium may be gained from the study of an isolated $\mathrm{Ce}$-$\mathrm{In}$ layer. Here we report on such an analysis, based on the generalized Slater-Koster (SK) approach.\cite{TakegaharaJPhysC1980}

\begin{figure}
\centering\includegraphics[width=0.8\linewidth]{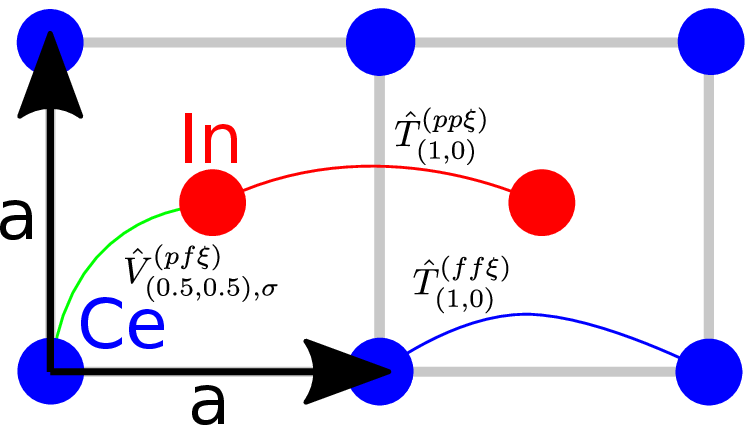}
\caption{Schematic representation of the Ce-In plane. The lines represent selected hopping and hybridization matrices $\hat{T}^{(pp\xi)}_{\delta r}$, $\hat{T}^{(ff\xi)}_{\delta r}$, and $\hat{V}^{(pf\xi)}_{\delta r' , \sigma}$ (cf. Hamiltonian Eq.~\ref{eq:slaterkosterhamiltonian}). Black arrows mark lattice translation vectors, whereas $a\approx 4.67$~\AA~is lattice spacing.}
\label{fig:cein_plane}
\end{figure}

We employ the effective model of Ce-In plane encompassing two intertwining square lattices of Ce and In ions, as is shown in~\hyperref[fig:cein_plane]{Fig.~\ref*{fig:cein_plane}}. As a result of the spin-orbit coupling, cerium $4f$ states split off into $j = 5/2$ and $j=7/2$ multiplets, with $j$ being total angular momentum. Following Ref.~\cite{maehira}, we retain only the $j=5/2$ sextet and indium $p_{x/y}$ states, which result in a five-orbital model involving $|p_a\rangle$, $|p_b\rangle$, $|f_a\rangle$, $|f_b\rangle$, and $|f_c\rangle$ states, defined as follows:
$$|p_{a\uparrow}\rangle = -|m=1, s=\uparrow\rangle,\quad\quad\quad |p_{a\downarrow}\rangle = |m=-1, s=\downarrow\rangle, $$
$$|p_{b\uparrow}\rangle = -|m=-1, s=\uparrow\rangle,\quad\quad\quad   |p_{b\downarrow}\rangle = |m=1, s=\downarrow\rangle, $$
$$|f_{a\uparrow}\rangle = \left|j=\frac{5}{2}, j^z=-\frac{5}{2}\right>,\quad\quad\quad |f_{a\downarrow}\rangle = \left|j=\frac{5}{2}, j^z=\frac{5}{2}\right>,$$
$$|f_{b\uparrow}\rangle = \left|j=\frac{5}{2}, j^z=-\frac{1}{2}\right>,\quad\quad\quad |f_{b\downarrow}\rangle = \left|j=\frac{5}{2}, j^z=\frac{1}{2}\right>, $$
$$ |f_{c\uparrow}\rangle = \left|j=\frac{5}{2}, j^z=\frac{3}{2}\right>,\quad\quad\quad |f_{c\downarrow}\rangle = \left|j=\frac{5}{2}, j^z=-\frac{3}{2}\right>.$$

Here $j^z$ is the $z$th component of the total angular momentum, whereas $s$ and $m$ denote its spin and orbital counterparts, respectively. In the case of $f_{a\sigma}$, $f_{b\sigma}$, and $f_{c\sigma}$ orbitals, the isospin index ($\sigma = \uparrow, \downarrow$) arises due to Kramers degeneracy.

The general TB Hamiltonian, governing the $p$-$f$ electron sector, takes the following form:
 
  \begin{widetext}
\begin{equation}
\begin{array}{l}
\label{eq:slaterkosterhamiltonian}
\hat{\mathcal{H}} =  \sum\limits_{\substack{\xi = \sigma, \pi  \mathbf{r}, \boldsymbol{\delta}\mathbf{r}, \sigma}} (pp\xi) \cdot \hat{p}_{\mathbf{r}+\boldsymbol{\delta}\mathbf{r}, \sigma}^\dagger \hat{T}_{\boldsymbol{\delta}\mathbf{r}}^{(pp\xi)} \hat{p}_{\mathbf{r}, \sigma} +      \sum\limits_{\substack{\xi = \sigma, \pi, \delta, \phi \\ \mathbf{r}, \boldsymbol{\delta}\mathbf{r}, \sigma}} (ff\xi) \cdot \hat{f}_{\mathbf{R}+\boldsymbol{\delta}\mathbf{r}, \sigma}^\dagger \hat{T}_{\boldsymbol{\delta}\mathbf{r}}^{(ff\xi)} \hat{f}_{\mathbf{R}, \sigma} +  \\   \left( \sum\limits_{\substack{\xi = \sigma, \pi \\ \mathbf{r}, \boldsymbol{\delta}\mathbf{r}, \sigma}} (pf\xi) \cdot \hat{p}_{\mathbf{R}+\boldsymbol{\delta}\mathbf{r}^\prime, \sigma}^\dagger \hat{V}_{\boldsymbol{\delta}\mathbf{r}^\prime, \sigma}^{(pf\xi)} \hat{f}_{\mathbf{R}, \sigma} + \mathrm{H.c.} \right) +  \sum\limits_{\mathbf{r}, \sigma}\Delta_p \cdot \hat{p}_{\mathbf{r}, \sigma}^\dagger \hat{p}_{\mathbf{r}, \sigma},\end{array}
\end{equation}
 \end{widetext}
where $(pp\sigma)$, $(pp\pi)$, $(ff\sigma)$, $(ff\pi)$, $(ff\delta)$, $(ff\phi)$, $(pf\sigma)$, and $(pf\pi)$ are respective SK integrals~\citep{slater}. The matrices $\hat{T}_{\boldsymbol{\delta}\mathbf{r}}^{(pp\xi)}$ and $ \hat{T}_{\boldsymbol{\delta}\mathbf{r}}^{(ff\xi)}$ describe hopping of $p$ and $f$ electrons, respectively, whereas  $\hat{V}_{\boldsymbol{\delta}\mathbf{r}^\prime, \sigma}^{(pf\xi)}$ is isospin-dependent $p$-$f$-electron hybridization. The positions $\mathbf{r}$ and $\mathbf{R}$ run over indium and cerium sites, respectively, whereas $\boldsymbol{\delta} \mathbf{r} = \pm (1, 0), \pm (0, 1)$, $\boldsymbol{\delta} \mathbf{r}^\prime = \pm (0.5, 0.5), \pm (0.5, 0.5)$ are vectors connecting neighboring atom positions in lattice units. Finally, $\Delta_p$ is the $p$-electron atomic level shift with respect to $4f$-level position. In the following we retain only nearest-neighbor hopping integrals and hybridization terms.

By employing the generalized SK approach~\citep{TakegaharaJPhysC1980}, we obtain the respective nearest-neighbor hopping and hybridization matrices as follows:

\begin{equation*}
 \hat{T}^{(pp\sigma)}_{(1,0)} = \hat{T}^{(pp\pi)}_{(0, 1)} =
\begin{pmatrix}
    0.5 & -0.5 \\
   -0.5 & 0.5
\end{pmatrix}
\end{equation*}

\begin{equation*}
 \hat{T}^{(pp\sigma)}_{(0, 1)} =  \hat{T}^{(pp\pi)}_{(1, 0)} =
\begin{pmatrix}
   0.5 & 0.5 \\
  0.5 & 0.5
\end{pmatrix}
\end{equation*}


\begin{equation*}
 \hat{T}^{(ff\sigma)}_{(1, 0)} = 
\begin{pmatrix}
     0.26785714 & -0.16940773 & 0.11978936 \\
     -0.16940773 & 0.10714286 & -0.07576144 \\
     0.11978936 & -0.07576144 & 0.05357143
\end{pmatrix} 
\end{equation*}
\begin{equation*}
  \hat{T}^{(ff\sigma)}_{(0, 1)} = 
\begin{pmatrix}
     0.26785714 & 0.16940773 & 0.11978936 \\
     0.16940773 & 0.10714286 & 0.07576144 \\
     0.11978936 & 0.07576144 & 0.05357143\\
\end{pmatrix}
 \end{equation*}
\begin{equation*}
 \hat{T}^{(ff\pi)}_{(1, 0)} = 
\begin{pmatrix}
    0.44642857 & -0.16940773 & 0.03992979 \\
     -0.16940773 & 0.17857143 & -0.17677670 \\
     0.03992979 & -0.17677670 & 0.23214286
\end{pmatrix}
\end{equation*}
\begin{equation*}
\hat{T}^{(ff\pi)}_{(0, 1)} =
 \begin{pmatrix}
 	0.44642857 & 0.16940773 & 0.03992979 \\
     0.16940773 & 0.17857143 & 0.17677670 \\
    0.03992979 & 0.17677670 & 0.23214286
 \end{pmatrix}
 \end{equation*}

\begin{equation*}
	\hat{T}^{(ff\delta)}_{(1, 0)} = 
	\begin{pmatrix}
 	    0.23214286 & 0.16940773 & -0.27950850 \\
 	    0.16940773 & 0.17857143 & -0.12626907 \\
 	    -0.27950850 & -0.12626907 & 0.44642857
 	\end{pmatrix}
	\end{equation*}
	\begin{equation*}
	\hat{T}^{(ff\delta)}_{(0, 1)} = 
	\begin{pmatrix}
 	    0.23214286 & -0.16940773 & -0.27950850 \\
 	    -0.16940773 & 0.17857143 & 0.12626907 \\
 	    -0.27950850 & 0.12626907 & 0.44642857
	\end{pmatrix}
\end{equation*}
\begin{equation*}
\hat{T}^{(ff\phi)}_{(1, 0)} = 
\begin{pmatrix}
     0.05357143 & 0.16940773 & 0.11978936 \\
     0.16940773 & 0.53571429 & 0.37880720 \\
     0.11978936 & 0.37880720 & 0.26785714
\end{pmatrix}
\end{equation*}
\begin{equation*}
 \hat{T}^{(ff\phi)}_{(0, 1)} = 
  \begin{pmatrix}
 	0.05357143 & -0.16940773 & 0.11978936 \\
    -0.16940773 & 0.53571429 & -0.37880720 \\
    0.11978936 & -0.37880720 & 0.26785714
 \end{pmatrix}
 \end{equation*}%
 
   \begin{widetext}
\begin{equation}
\begin{array}{l}
\label{eq:vpf_sigma}
\hat{V}^{(pf\sigma)\dagger}_{(0.5, 0.5), \sigma} = \hat{V}^{(pf\sigma)\dagger}_{(-0.5,-0.5), \sigma} =
 \begin{pmatrix}
      0.36596253 & i \sigma 0.36596253 \\
      i \sigma 0.23145502 & -0.23145502 \\
     -0.16366342 & - i \sigma 0.16366342
 \end{pmatrix}, \\
 \hat{V}^{(pf\sigma)\dagger}_{(-0.5, 0.5), \sigma} =  \hat{V}^{(pf\sigma)\dagger}_{(0.5, -0.5), \sigma} = 
\begin{pmatrix}
     0.36596253 & -i \sigma 0.36596253 \\
     -i \sigma 0.23145502  & -0.23145502 \\
     -0.16366342 & i \sigma 0.16366342
 \end{pmatrix}
 \end{array}
 \end{equation}

\begin{equation} 
\begin{array}{l}
 \label{eq:vpf_pi}
   \hat{V}^{(pf\pi)\dagger}_{(0, 0), \sigma} = \hat{V}^{(pf\pi)\dagger}_{(-1, -1), \sigma} =
  \begin{pmatrix}
     -0.44821073 & i\sigma 0.44821073 \\
     -i\sigma 0.09449112 & -0.09449112 \\
     -0.06681531 & i\sigma 0.06681531
   \end{pmatrix},\\
  \hat{V}^{(pf\pi)\dagger}_{(-1, 0), \sigma} = \hat{V}^{(pf\pi)\dagger}_{(0, -1), \sigma} =
   \begin{pmatrix}
     -0.44821073 & -i\sigma 0.44821073 \\
     i\sigma 0.09449112 & -0.09449112 \\
     -0.06681531  & -i\sigma 0.06681531
   \end{pmatrix}
   \end{array}
   \end{equation}
      \end{widetext}

Note that the hopping and hybridization matrices involving $f$ electrons ($\hat{T}_{\boldsymbol{\delta}\mathbf{r}, \sigma}^{(ff\xi)}$ and $\hat{T}_{\boldsymbol{\delta}\mathbf{r}, \sigma}^{(pf\xi)}$) take a nonstandard form due to spin-orbit coupling that mixes the orbitals through the Clebsch-Gordan coefficients. If only $\sigma$-bonds are retained, the above model reduces to that considered in Ref.~\cite{TakegaharaJPhysC1980}. As is detailed below, we find that $\sigma$ bonds are not sufficient to obtain a good quantitative agreement with the ARPES data simultaneously around  \={X} and  \={M} points in the Brillouin zone. Minimally, $\pi$ bonds for the dispersive $p$ bands should be incorporated to match the data on an overall scale.

\subsection{The procedure of fitting to ARPES data}
To relate our TB model Eq.~\eqref{eq:slaterkosterhamiltonian} to ARPES data, we set $\Delta_p \equiv 0.1\,\mathrm{eV}$ and treat $(pp\sigma)$, $(pp\pi)$, $(ff\sigma)$, $(pf\sigma)$, as well as the chemical potential $\mu$ as fitting parameters, whereas the remaining SK integrals are set to zero. The value of $\Delta_p$ has been fixed in order to reduce the number of characteristic energies that need to be accurately extracted from the data (below, we argue that this is a reasonable choice by performing an alternative fit for a different parameter configuration). The remaining parameters are determined from the five conditions based on ARPES measurements: (\emph{i}) the lowest TB band energy at the \={X} point is $-0.586\,\mathrm{eV}$, (\emph{ii}) the second-lowest TB band energy at the \={M} point is $-0.469\,\mathrm{eV}$, (\emph{iii}) the Fermi wave vector of the lowest TB band along the  $\bar{\Gamma}$-\={X} direction reads $(0.35 \frac{\pi}{a}, 0)$ (with $a \approx  4.67\,\text{\AA}$ being lattice spacing), (\emph{iv}) the Fermi wave vector of the lowest TB band along the  \={X}-\={M} direction is $(0.320418\,\text{\AA}^{-1}, \frac{\pi}{a})$, and (\emph{v})  $4f$ occupancy is selected as one per $\mathrm{Ce}$ ion. The computations have been carried out for $400 \times 400$ lattice and temperature $T = 0.01\,\mathrm{eV}/k_B$ with the absolute tolerance for the Brillouin-zone sums set to no less then than $10^{-10}$. The system of five-equations was solved using SciPy \verb!hybr! method, with tolerance taken as $10^{-10}$. The obtained parameter values are summarized in~\hyperref[tab:parameters]{Table~\ref*{tab:parameters}}, and the corresponding orbital occupancies are provided in~\hyperref[tab:occ1]{Table~\ref*{tab:occ1}}.

\subsection{Discussion}
As is apparent from~\hyperref[tab:parameters]{Table~\ref*{tab:parameters}}, the $(pp\sigma)$ parameter is dominant, whereas we obtain much smaller value of direct $f$-$f$ hopping (controlled by $(ff\sigma)$), which indicates a nearly-localized nature of $f$-electron states. This observation justifies \emph{a posteriori} neglecting higher-order SK parameter in the $f$-electron sector [$(ff\pi)$, $(ff\delta)$, and $(ff\phi)$] as those would acquire even smaller values. Also, because of that factor, we expect a minor deviation of $n_f$ from unity and hence the assumption $n_f = 1$ should be irrelevant to the discussion of electronic structure. The $(pp\pi)$ integral value turns out to be an order of magnitude smaller than that of $(pp\sigma)$, with $|(pp\pi)| \sim |(ff\sigma)|$. The $(pp\sigma)$ and $(pp\pi)$ integrals have opposite signs, which is consistent with the Harrison theory prediction based on muffin-tin pseudopotentials~\citep{HarrisonBook}. The latter approach yields, however, substantially larger $(pp\pi) = -0.5 (pp\sigma) \approx -0.163499$. Such discrepancies are expected for real materials. The calculated band structure is displayed and compared with the ARPES spectra in~\hyperref[fig:czworka]{Fig.~\ref*{fig:czworka}} of the main text. The structure reasonable matches the experiment in the vicinity of both \={M} and \={X} points.

\begin{figure}
\centering\includegraphics[width=0.95\linewidth]{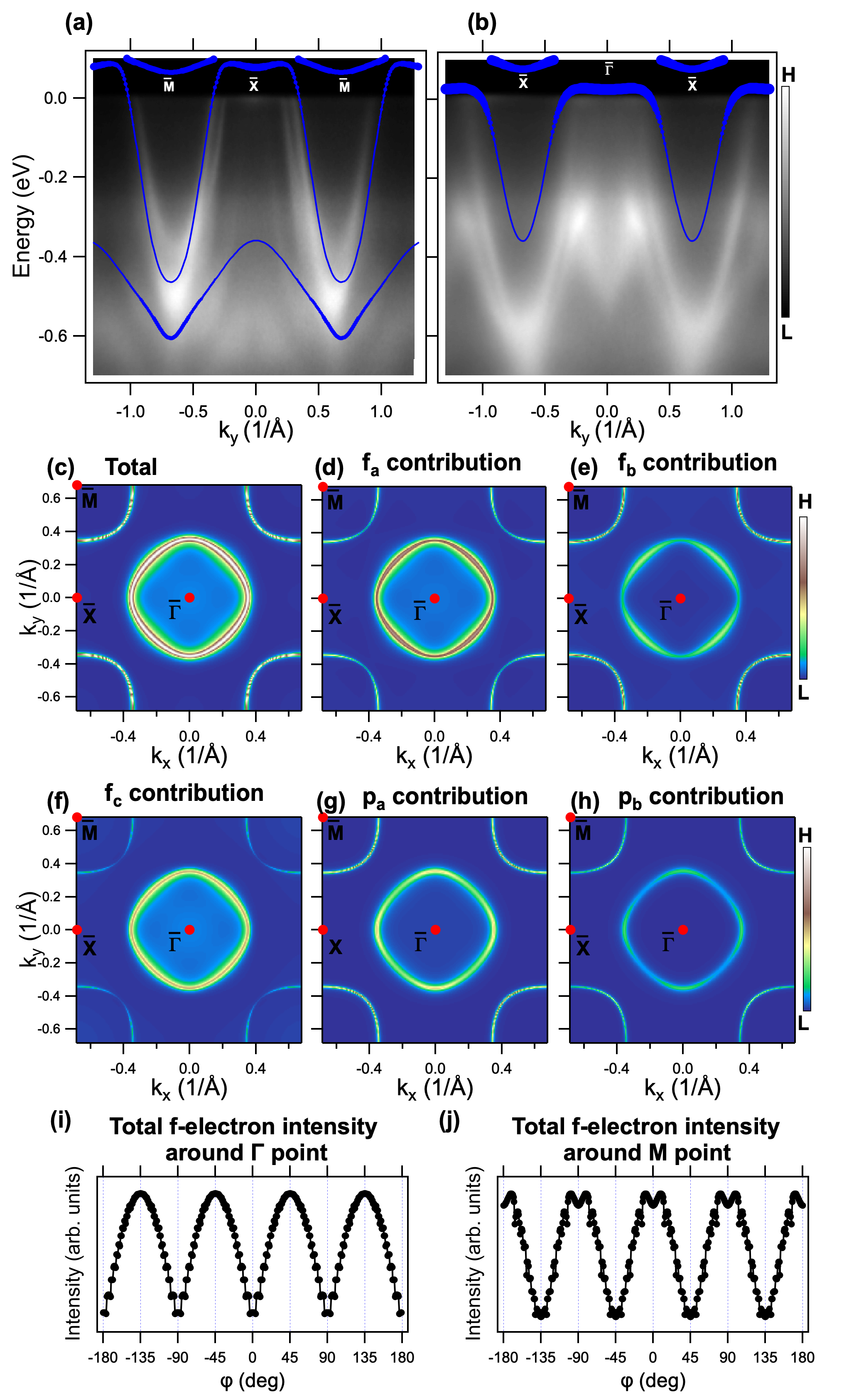}
\caption{Theoretical band structure for Ce-In planes (solid lines) obtained by tight-binding (TB) calculations. TB parameters have been adjusted by fitting theoretical bands to ARPES data with assumption $(pp\pi)=0$. This is the alternative version of TBA  fit model presented in main part of the paper. The comparison of fitted model with the spectra collected along \={M}-\={X}-\={M} (a) and \={X}-$\bar{\Gamma}$-\={X} direction is shown. The marker size is proportional to the contribution of $f$ orbital to the band. Panel (c) shows the distribution of the spectral weight on the Fermi surface. The projections on different base states are shown in (d)-(h). The Lorentzian broadening of 15~meV was applied in order to plot such maps. Panels (i) and (j) show total $f$-electron intensity extracted from TBA calculations along FS contours around $\Gamma$ and M points. Azimuthal angle $\phi$=0 corresponds to $\bar{\Gamma}$-\={X} direction in (i) and \={M}-\={X} direction in (j).}
\label{fig:arpesdata}
\end{figure}

To illustrate the relevance of $\pi$ bonds, we have also performed an analogous fit for $(pp\pi) \equiv 0$. Now, since we have one less free parameter, the value of the $p$-orbital level $\Delta_p$ may be determined directly from the ARPES data. The fitting conditions are now: (\textit{i}) energy of the lowest TB band at the \={M} point is $-0.623\,\mathrm{eV}$, (\textit{ii}) energy of the second-lowest TB band at the \={M} point is $-0.469\,\mathrm{eV}$, (\textit{iii}) energy of the third-lowest band at the \={M} point is $0.03\,\mathrm{eV}$, (\textit{iv}) the Fermi wave vector of the lowest TB band along the \={X}-\={M} direction is $(0.320418 \, \text{\AA}^{-1},  \frac{\pi}{a})$, and (\textit{v}) the  total $4f$-electron occupancy is one per $\mathrm{Ce}$ ion. Temperature is taken as $T = 0.01\,\mathrm{eV}/k_B$ and other numerical details remain same as above. The condition (\textit{iii}) ensures that the electronlike band with substantial $f$-electron contribution remains above the Fermi level and no spurious electron-pockets appear near the \={M} point. The resultant model parameters are listed in~\hyperref[tab:parameters2]{Table~\ref*{tab:parameters2}} and orbital occupancies in~\hyperref[tab:occupation2]{Table~\ref*{tab:occupation2}}. Note that when $(pp\pi) \equiv 0$ is imposed, a much larger value of direct $f$-$f$-electron hopping $(ff\sigma) \approx 0.10$ comes out from the fitting procedure. In this situation, neglecting $(ff\pi)$, $(ff\delta)$, and $(ff\phi)$ is no longer justified. Note that both the sign of $\Delta_p$ and its order of magnitude are the same as those taken for the previous fit. The calculated band structure for the $(pp\pi) \equiv 0$ case is compared with the ARPES data in~\hyperref[fig:arpesdata]{Fig.~\ref*{fig:arpesdata}}. Whereas the overlap of calculated and measured band structures remains satisfactory near the \={M} point, substantial discrepancies are seen in the vicinity of the \={X} point. The calculated Fermi surfaces are similar for both zero- and nonzero value of $(pp\pi)$, which points towards a degree of universality of the low-energy electronic structure for the Ce-In termination. 

\begin{table}
\caption{Values of the model parameters obtained on the basis of ARPES spectra for CeCoIn$_5$, with nonzero Slater-Koster integral $(pp\pi)$.}
\centering
\begin{tabular}{cl}
Parameter & Value (eV) \\
\hline
$(pp\sigma)$ & 0.3269978375 \\
$(pp\pi)$ & -0.0378542939 \\
$(ff\sigma)$ & 0.0479663350\\
$(pf\sigma)$ & 0.1135277632\\
$\mu$ & -0.0349293094\\
$\delta_p$ & 0.1\\
\hline
\label{tab:parameters}
\end{tabular}
\end{table}

\begin{table}
\centering
\caption{Orbital-resolved occupancies for the five-orbital model with nonzero value of Slater-Koster integral $(pp\pi)$.}
\begin{tabular}{cl}
Orbital & Occupancy \\
\hline
$p_a$ &  0.8912382550\\
$p_b$ &   0.8337279039\\
$f_a$ & 0.4923705187\\
$f_b$ &  0.2433943023\\
$f_c$ & 0.2642351790\\
\hline
\label{tab:occ1}
\end{tabular}
\end{table}

\begin{table}
\centering
\caption{Summary of the model parameters obtained based on ARPES spectra for CeCoIn$_5$. The case of $(pp\pi)\equiv 0$.}
\begin{tabular}{cl}
Parameter & Value (eV) \\
\hline
$(pp\sigma)$ & 0.3550378608 \\
$(pp\pi)$ & 0  \\
$(ff\sigma)$ & 0.1021043903\\
$(pf\sigma)$ & 0.2621072652 \\
$\mu$ & -0.1118485020\\
$\delta_p$ & 0.2388661234\\
\hline
\label{tab:parameters2}
\end{tabular}
\end{table}

\begin{table}
\centering
\caption{Orbital occupancies for the five-orbital model of CeIn plane. The case of $(pp\pi)\equiv 0$.}
\begin{tabular}{cl}
Orbital & Occupancy \\
\hline
$p_a$ &  0.7436812923\\
$p_b$ &  0.6577470862 \\
$f_a$ & 0.5741199928\\
$f_b$ &  0.2310639249\\
$f_c$ & 0.1948160823\\
\hline
\label{tab:occupation2}
\end{tabular}
\end{table}

\subsection{Anisotropy of the hybridization matrix in k-space}
Experimental data (\hyperref[fig:arpesdata]{Fig.~\ref*{fig:arpesdata}}), first-principle (\hyperref[fig:bsf]{Fig.~\ref*{fig:bsf}}) and TB (\hyperref[fig:arpesdata]{Fig.~\ref*{fig:arpesdata}}) analysis consistently point towards strongly anisotropic spectral weight distribution of the $f$ electrons along the FS, particularly in the vicinity of the $\Gamma$ point. It is instructive to provide an interpretation of those anisotropic features in terms of orbital-dependent hybridization between $f$ and $p$ orbitals, as obtained within the SK framework. This discussion is at most qualitative, since the matrix-element contribution to measured intensities is entirely neglected.

\begin{figure}
\centering\includegraphics[width=0.9\linewidth]{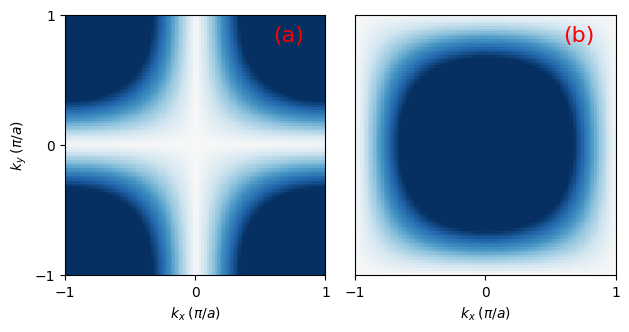}
\caption{Schematic representation of the $\mathbf{k}$-space hybridization matrix element magnitudes, $|| \hat{V}^{(pf\sigma)\dagger}_{\mathbf{k}\sigma}|_{\alpha\beta} ||$ (cf. Eq.~\eqref{eq:vpf_sigma}). Panel (a) shows directional characteristics of $||\hat{V}^{(pf\sigma)\dagger}_{\mathbf{k}\sigma}|_{12}||$, $||\hat{V}^{(pf\sigma)\dagger}_{\mathbf{k}\sigma}|_{21}||$, and $||\hat{V}^{(pf\sigma)\dagger}_{\mathbf{k}\sigma}|_{32}||$, whereas panel (b) those of $||\hat{V}^{(pf\sigma)\dagger}_{\mathbf{k}\sigma}|_{11}||$, $||\hat{V}^{(pf\sigma)\dagger}_{\mathbf{k}\sigma}|_{22}||$, and $||\hat{V}^{(pf\sigma)\dagger}_{\mathbf{k}\sigma}|_{31}||$. Blue color intensity represents nonzero hybridization magnitude. Note the nodes appearing along $\bar{\Gamma}$-\={X}  direction for the case (a).}
\label{fig:hybridization}
\end{figure}

The structure of the dominant ($\hat{V}^{(pf\sigma)}$) hybridization matrix indeed exhibits direction-dependent behavior as can be seen from~\eqref{eq:vpf_sigma}; the sign of every-second $\hat{V}^{(pf\sigma)}$ entry alternates after rotations by $90^\circ$. In the wave-vector-space representation, this results in the appearance of nodes for half of the hybridization matrix elements, whereas the remaining ones are nodeless. This is illustrated in~\hyperref[fig:hybridization]{Fig.~\ref*{fig:hybridization}}. As an overall effect, an increased $f$-electron contribution along diagonal ($\bar{\Gamma}$-\={M}) close to the $\bar{\Gamma}$ point is expected as is confirmed by direct numerical calculation (cf.~\hyperref[fig:arpesdata]{Fig.~\ref*{fig:arpesdata}}).

\end{document}